# EXPERIENCES WITH THE FERMILAB HINS 325 MHZ RFQ

R. C. Webber#, T. Khabiboulline, R. Madrak, G. Romanov, V. Scarpine, J. Steimel, D. Wildman
FNAL*, Batavia, IL 60510, U.S.A.

*Abstract*

The Fermilab High Intensity Neutrino Source program has built and commissioned a pulsed 325 MHz RFQ. The RFQ has successfully accelerated a proton beam at the design RF power. Experiences encountered during RFQ conditioning, including the symptoms and cause of a run-away detuning problem, and the first beam results are reported.

## INTRODUCTION

A Fermilab/Argonne collaboration specified a pulsed 2.5 MeV, 325 MHz RFQ to accelerate H$^-$ ions for the Fermilab High Intensity Neutrino Source (HINS) program [1][2][3][4]. ACCSYS Technology Inc., Stockton, CA, fabricated the RFQ and high power RF conditioning began at Fermilab in December 2008. Anomalous detuning of the RFQ resonance became manifest as the average applied RF power increased. Accelerated beam was achieved in January 2010.

## RF CONDITIONING

RF conditioning of the RFQ began in December 2008 with the goal of establishing stable RF operation at ~110% of the design power. During the process, several mechanical problems appeared that required repair. Overall, conditioning was a slow process requiring many tens of hours of closely attended operation.

### Input Coupler Issues

First attempts at high power operation were limited by excessive sparking in the input couplers. Both couplers were removed and, in one, a vacuum leak was discovered at a soldered vacuum joint. Inspection revealed considerable erosion on the coupling loop surfaces and sputtered solder from the loop connection joints. Some erosion and sputtered solder was also seen on the RFQ vanes near the coupler locations. The vacuum leak was repaired and the loops reattached with copper welds.

### Detuning Problems

As RF conditioning proceeded and the applied average power was increased, an anomalous and unstable detuning of the structure appeared [5]. A run-away detuning condition would set in at ~250 watts average RF power, less than 10% of design. An increase in sparking rate accompanied this problem.

Figure 1 shows the course of a detuning cycle. The detuning could exceed several hundred kilohertz if RF power was reduced sufficiently to prevent vacuum interlock trips. Once the severely detuned state was established, only tens of watts average RF power was necessary to maintain it. The effect was completely reversible; after twenty minutes without RF power, the RFQ returned to 'normal'. The time-to-onset of the run-away condition correlated strongly with the average applied RF power.

Figure 2 shows the mode spectrum of the RFQ changing dynamically during run-away. A parasitic mode moves lower in frequency over time, eventually splitting the desired 325 MHz accelerating mode (leftmost peak).

Simulations were performed to investigate possible causes. Different mechanical problems with tuners, couplers, seals etc. were investigated, but none showed the obvious resonance characteristics that were observed.

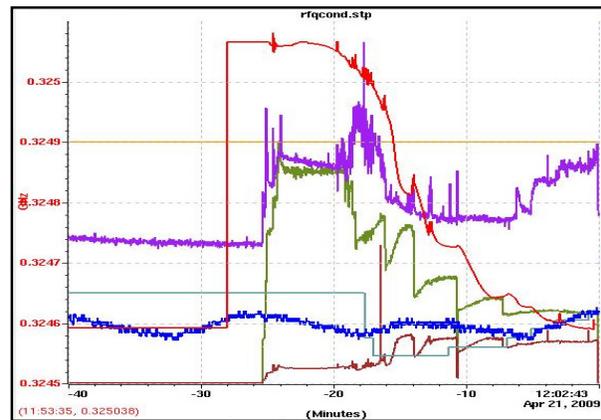

Figure 1. RFQ frequency (red, 600 kHz span), peak power (green, 600 kW span), and pulse width (cyan, 4 ms span) vs. time (40 min span) at constant 2 Hz pulse rate

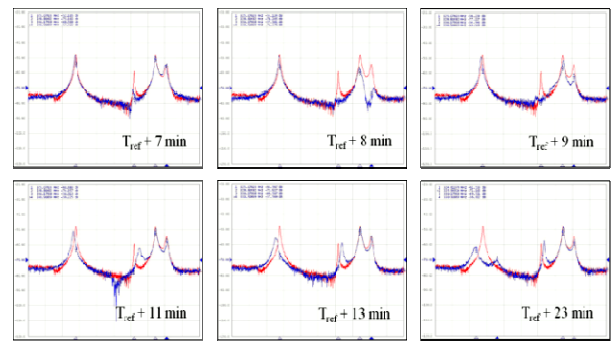

Figure 2. Network analyzer plot of 'normal' RFQ mode spectrum (red) and mode spectra (blue) as RFQ slips into run-away state. Scale is 10 db/div and 11 MHz full span. $T_{ref}$ is time from RF turn-on at ambient temperature.

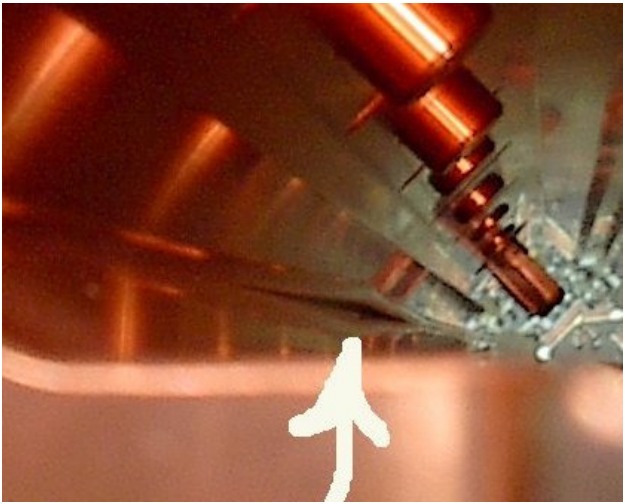

Figure 3. RFQ End-view and bulging RF c-seal responsible for run-away detuning

When the RFQ was opened for inspection, a C-seal, designed to serve as an RF current by-pass across the longitudinal joint between quadrants, was found dislocated from its groove. A 12-18 in. section at one joint, about 18 in. from the RFQ high-energy end, was bulging into the RF volume. It was discoloured from heat and sparking. Figure 3 shows the bad C-seal.

The bulging seal was added to the CST MicroWave Studio solid model of RFQ. Simulations indicated a parasitic mode corresponding to a resonator almost exactly λ/2 at the RFQ operating frequency and strongly coupled to the desired quadrupole field. As the C-seal heats and bulges farther from the groove, the parasitic frequency moves ultimately crossing the quadrupole mode. The well-known mode-mixing picture in Figure 4 is the result.

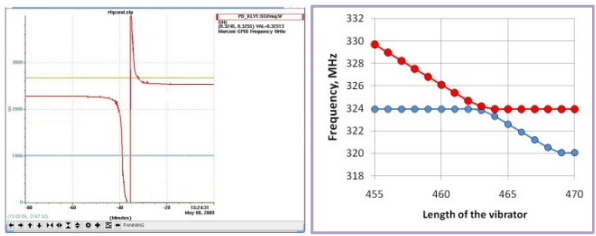

Figure 4. Mode mixing in RFQ, experiment (left) and simulation (right)

ACCSYS replaced both C-seals in the affected quadrant at their facility. Subsequently, the RFQ has been powered to well beyond the level previously sufficient to cause the run-away and the problem appears to be solved.

### Water Leaks

In spite of three Varian TV-1001 900 l/s vacuum turbo pumps on the RFQ vacuum vessel, the pressure persistently showed notable activity during RF operation and high sensitivity to ambient temperature. A residual gas analyzer showed water as the primary contaminant. The water source was attributed to permeation [6] through Viton o-rings sealing some thirty-four water tube connections within the vacuum vessel.

## FIRST BEAM TESTS

### Configuration

An ion source and low energy transport (LEBT) line with two solenoid magnets were configured to inject a 50 keV proton beam into the RFQ. In this set-up, one beam current transformer was the only beam diagnostic between the source and the RFQ.

A 2.3m diagnostic line [7] comprising a beam current transformer, three transverse wire scanners, two button BPMS, and a beam absorber was installed downstream of the RFQ. No focusing elements were included in the diagnostic line.

### Results

There was evidence of beam from the RFQ the first time that the ion source and the RFQ pulses were made coincident. After some steering in the LEBT and tuning of the solenoids, 4 mA of beam was obtained from the RFQ with about 14 mA observed in the LEBT. Earlier measurements [8] had suggested a high fraction of $H_2^+$ and $H_3^+$ ions, perhaps ≥70%, in the source beam. Due to this, a useful measure of RFQ efficiency was not possible.

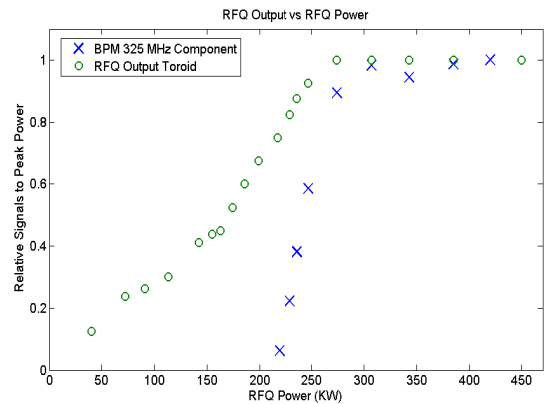

Figure 5. Relative transmission of total beam current (green circle) and bunched, accelerated beam current (blue x) through RFQ as function of RF power

The relative beam transmission through the RFQ was measured as a function of RF power. Downstream of the RFQ, signals were taken from the beam current transformer for a measure of total beam and from a BPM through a 325 MHZ bandpass filter for a measure of the bunched, presumably accelerated, beam. Figure 5 shows the results. The acceleration knee is found to be just below 300 kW, near the 350 kW level expected from the design. Below that power level, the accelerated beam drops quickly to nothing at 210 kW. The total transmitted beam drops slowly and disappears only as the RF power approaches zero. This indicates that the RFQ serves as a viable focusing and transport channel for non-accelerated particles.

Two BPM pick-ups in the diagnostic line provided signals for a time-of-flight energy measurement. Figure 6 shows these signals. The BPMs are separated by 0.96 m. The corresponding transit time delay for 2.50 MeV protons is 43.96 ns with a sensitivity of 8.7 ps/keV or 1°/keV (325 MHz phase). The phase between the two signals could be determined with great accuracy; however, the integer number of RF cycles in the delay can be ambiguous without a sharp temporal edge on the beam. The most likely delay was determined to be 14 RF cycles + 940ps = 44.012 ns. This indicates the beam energy ~7 keV less than the 2.50 MeV design.

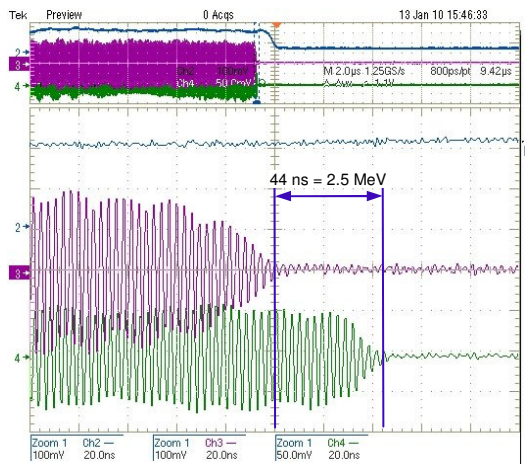

Figure 6. Time of flight energy measurement (20 ns/div)

## PRESENT STATUS

Following the successful 2.5 MeV beam test, the RFQ vacuum tank was opened to replace the 34 Viton o-ring water seals with metal seals. The low-energy copper end-wall of the RFQ showed signs of impinging radiation in a circle, the size of the upstream beam tube, around the beam aperture. Presumably, this is mostly due to unwanted, unfocused ion species from the source.

The RFQ vane tips showed some erosion and pitting for about 30 cm beginning at the low-energy end. This is a cause for concern. Several theories are posited as a cause, including: a) heating or spallation due to poorly matched or undesired ions impacting the surfaces and b) excessive water molecules in the vacuum acting alone or in conjunction with the beam to induce corona and sparking. Subsequent investigation of the ion source did confirm that only ~30% of the output current is protons. Studies are underway to determine a better operating point for the source; simultaneously, work is beginning to configure an H- source to mate to the RFQ.

When checking the integrity of the newly installed metal water tubing seals, other water channel to vacuum leaks were discovered. These leaks appear to be where fabrication machining operations required boring holes into the copper structure that were later plugged. The plugged regions are not all leak tight and some are inaccessible without full disassembly of the RFQ; options are now being investigated.

## FUTURE HINS RFQ BEAM OPERATIONS

The RFQ will be reinstalled in the HINS beam line in autumn 2010. New diagnostics will be in place to characterize fully the transverse and longitudinal properties of the output beam. Within the coming year, an H- ion source will replace the present proton source.

This system will serve as a low-energy beam facility in support of future Fermilab projects, including the proposed Project X. It will provide a platform for testing and development of new beam instrumentation techniques and it will serve as a valuable hands-on training facility for future accelerator scientists.

## SUMMARY

The HINS RFQ has successfully accelerated protons to 2.5 MeV, despite run-away detuning problems, RF input coupler vacuum leaks, and water-to-vacuum leaks. Operation will resume later in 2010 to provide beam in support of future proton/H- accelerator missions.